\documentclass{PoS}
\usepackage{amsmath}
\usepackage{amssymb}
\usepackage{graphicx}
\usepackage{epsfig}
\usepackage{color}
\usepackage{latexsym}
\usepackage{amsmath}
\usepackage{amssymb}
\usepackage{dsfont}
\usepackage{verbatim}

\title{Charmonium-like states from scattering on the lattice}

\ShortTitle{Charmonium-like states from scattering on the lattice}

\author{\speaker{Sasa Prelovsek}\\
       Faculty of Mathematics and Physics, University of Ljubljana, Jadranska 19, 1000 Ljubljana, Slovenia\\
       and Jozef Stefan Institute, Jamova 39, 1000 Ljubljana, Slovenia\\
       E-mail: \email{sasa.prelovsek@ijs.si}}

\author{Luka Leskovec\\
       Jozef Stefan Institute, Jamova 39, 1000 Ljubljana, Slovenia\\
       E-mail: \email{luka.leskovec@ijs.si}}

\author{Daniel Mohler\\
       Fermi National Accelerator Laboratory, Batavia, Illinois 60510-5011, USA\\
       E-mail: \email{dmohler@fnal.gov}}

\abstract{Three charmonium-like states were studied using lattice QCD. The candidate for $X(3872)$ was found slightly below $D\bar D^*$ threshold in the channel with $J^{PC}=1^{++}$ and $I=0$, where $\bar cc$ as well as  $D\bar{D}^{*}$ and $J/\psi \omega$ interpolating operators were used.  A charmonium-like channel with $J^{PC}=1^{+-}$ and $I=1$ was also studied, as the recently discovered $Z_{c}^{+}(3900)$ might reside there. Here $J/\psi \pi$ and $\bar{D}D^{*}$ scattering states were found, but no candidate for the $Z_{c}^{+}(3900)$. We present also preliminary results for the $J^{PC}=0^{++}$ charmonium channel, where $\bar cc$, $D\bar{D}$ and $J/\psi \omega$ interpolating operators were used.
A candidate for a resonance, $\chi_{c0}^{'}$, that couples to $D\bar{D}$ in $J^{PC}=0^{++}$ was found.}

\FullConference{31st International Symposium on Lattice Field Theory - LATTICE 2013\\
		July 29 - August 3, 2013\\
		Mainz, Germany}

\begin{document}

\section{Introduction}
For a long time the charmonium spectrum has been calculated via quark potential models offering certain predictions, but also missing some very important states, for example the $X(3872)$. However recently lattice QCD has become able to offer charmonium spectrum predictions from first principles. The most recent and most extensive lattice calculation of the charmonium spectrum was done by the Hadron Spectrum Collaboration within the so-called single-meson approach \cite{Liu:2012ze}. This approach ignores strong decays of resonances and neglects threshold effects, which can be sizable for near threshold states. These were incorporated by Bali et al. \cite{Bali:2011rd}, where in addition to $\bar{c}c$ interpolating operators also scattering operators were used. While a general view on the spectrum was gained, there still remained questions regarding certain states and channels.\\
In these proceeding we present our study of a small subset of the charmonium(-like) states, which are still not well understood. In addition to using simple $\bar{c}c$ interpolating operators, we also include meson-meson scattering interpolating operators in order to identify states near thresholds reliably.\\
After a brief overview of our simulations and analysis, we present our results for $X(3872)$ and $D\bar{D}^{*}$ scattering length \cite{Prelovsek:2013cra} in Sec III. Sec IV is dedicated to the exotic $Z_{c}^{+}(3900)$ state, which we search for in the channel with $J^{PC}=1^{+-}$ and $I=1$  \cite{Prelovsek:2013xba}. Preliminary results of the $J^{PC}=0^{++}$ channel are presented in Section IV, where we find a possible candidate for a resonance that couples to $D\bar{D}$ in $J^{PC}=0^{++}$.

Recent reviews of charmonium and open charm spectroscopy from lattice are given in \cite{Prelovsek:2013cta, Mohler:2012nh}.

\section{About the configurations and methods}
Our results are based on one ensemble of gauge configurations on a $16^3 \times 32$ lattice with the spacing being $a=0.1239(13)$ fm \cite{Hasenfratz:2008fg,Hasenfratz:2008ce}. It is a $N_f=2$ ensemble, with clover Wilson dynamical u and d quarks and valence u, d and c quarks. The u and d quarks are mass degenerate, giving our lattice an exact isospin symmetry with a pion mass of $m_{\pi}=266(4)$ MeV.\\
The charm quarks are treated using the Fermilab method \cite{ElKhadra:1996mp,Oktay:2008ex}, so we do not calculate absolute energies, but rather the energy differences between the state we are interested in and the spin averaged charmonium mass, $m_{sa}=(m_{\eta_{c}} + 3m_{J/\psi})/4$ with $am_{\eta_c}=1.47392(31)$ and $am_{J/\psi}=1.54171(43)$. For completeness we also provide $D$ and $D^{*}$ masses, $am_{D}=0.9801(10)$ and $am_{D^{*}}=1.0629(13)$.\\
We build a correlation matrix from distinct operators $O_{i}$ with desired quantum numbers, for each channel in the following sections. In the correlation matrix we include all Wick contractions, except those with a disconnected charm quark, as their effects on charmonium states is suppressed due to the OZI rule. To evaluate the Wick contractions we use the distillation method \cite{Peardon:2009gh}, which is an all-to-all method and allows us to calculate the needed light disconnected diagrams.\\
We obtain the physical states from the time evolution of the correlation matrix $C_{ij}=\langle 0 | O_{i}(t) O_{j}^{\dagger}(0) | 0 \rangle = \sum_{n}Z_{i}^{n}Z_{j}^{n*}e^{-E_{n}t}$, where $Z_{i}^{n}=\langle O_{i} | n \rangle$. By solving the generalized eigenvalue problem \cite{Blossier:2009kd} $C(t)v_{n}(t)=\lambda_{n}(t)C(t_{0})v_{n}(t)$, we obtain the eigenvalues $\lambda_{n}(t)$, which have a distinct time dependence, $\lambda_{n}(t) \propto e^{-E_{n}t}$, from where we can extract the physical state energy $E_{n}$.\\
The extracted energies belong to a variety of physical states that have the quantum numbers of the interpolating operators. The resulting spectrum contains discrete scattering states as well as bound states or resonances, which usually appear as additional energy levels.

\section{$X(3872)$}

\begin{figure}[hbt]
\vspace{0.3cm}
\centering
\includegraphics[width=0.8\textwidth]{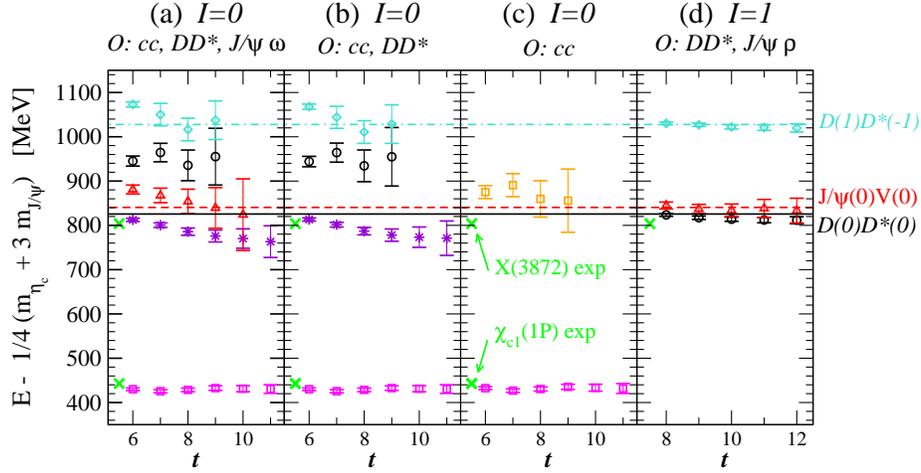} 
\caption{Energy levels for $I=0$ and $I=1$ in the $X(3872)$ channel.  Symbols represent $E_{n} - (m_{\eta_{c}} + 3m_{J/\psi})/4$ in the plateau region; $E_{n}$ are physical state energies. Dashed lines represent non-interacting $D\bar{D}^{*}$ and $J/\psi V$, ($V=\rho,\omega$), scattering levels, while the basis used is indicated above each graph respectively.}
\label{Xfig}
\end{figure}

While the charmonium state $X(3872)$ has been known for a long time, its quantum numbers have only recently been determined to be $J^{PC}=1^{++}$ \cite{Aaij:2013zoa}. Experimentally it lies within $1$ MeV of the $D^{0}\bar{D}^{*0}$ threshold and decays both into the isoscalar channel $J/\psi \omega$ as well as into the isovector channel $J/\psi \rho$. We simulate both $I=0$ and $I=1$ channels using 8 $\bar{c}c$ (only for $I=0$), 3 $D\bar{D}^{*}$ and 2 $J/\psi V$ ($V=\rho,\omega$) interpolating operators listed in \cite{Prelovsek:2013cra}. The energy levels are presented in Fig. \ref{Xfig}.

$I=1$ results are easiest to understand, as we find only scattering levels $D(0)\bar{D}^*(0)$, $J/\psi (0) \rho(0)$ and $D(1)\bar{D}^*(-1)$. They lie very near their respective non interacting energies, which indicates negligible interaction in this channel. We find no additional level and hence no candidate for $X(3872)$ in the $I=1$ channel, which may be attributed to exact isospin symmetry on our lattice.\\
In the $I=0$ channel we observe two additional states; the lowest (pink squares) is identified with the $\chi_{c1}(1P)$, while the higher one is attributed to $X(3872)$. By removing  $J/\psi \omega$ operators from the basis, as seen Fig. \ref{Xfig} (b), we find that the $J/\psi \omega$ is weakly coupled to the rest of the system and thus we remove it from further analysis. The two states, the purple stars and the black circles, have two possible identifications. The first scenario is that the purple stars correspond to a weakly bound $X(3872)$, which is slightly below the $D\bar{D}^*$ threshold, while the black circles correspond to an up shifted $D\bar{D}^*$ scattering state. In the second scenario the purple stars correspond to a downshifted $D\bar{D}^*$ scattering state and while black circles correspond to a resonance above the $D\bar{D}^*$ threshold.\\
Independently of the above interpretation we evaluate the scattering length of $D\bar{D}^*$ scattering. By extracting phase shifts for the second and third level of subplot (b) in Fig. \ref{Xfig}, and fitting the data points to the effective range expansion, $p \cot{\delta(p)}=\frac{1}{a_{0}^{DD^*}} + \frac{1}{2}r_{0}^{DD^*}p^2$, we get the scattering length for $D\bar{D}^*$ elastic scattering. For fitted energies and phase shifts cf. Table \ref{Xtab}.

\begin{table}[htb]
\centering
\begin{tabular}{c|ccc}
level $n$   & $E_n\!-\!\tfrac{1}{4}(m_{\eta_c}\!+\!3m_{J/\psi})$ [MeV ] & $p^2$  [GeV$^2$]  & $p\cdot \cot \delta(p)$[GeV] \\ 
\hline
2 & $785(8)$  & $-0.075(15)$ & $-0.21(5)$ \\
3 & $946(11)$ & $0.231(22)$ &  $0.17(9)$\\
\end{tabular}
\caption{Fitted energies for the second and third level from the $6\times 6$  $C_{ij}(t)$ based on $O^{\bar cc}_{1,3,5},~O^{DD^*}_{1,2,3}$. The $p$ denotes $D$ and $D^*$ momentum and $\delta(p)$ denotes their scattering phase shift. }
\label{Xtab}
\end{table}

According to \cite{Sasaki:2006jn}, if the scattering length is negative, then the first scenario is realized, and if the scattering length is positive, then the second scenario is realized. In our simulation we find the scattering length to be $a_{0}^{DD^*} = -1.7 \pm 0.4~$fm, and thus our simulation prefers the first scenario, where the $X(3872)$ is below threshold. Infinite volume extraction gives the mass of the $X(3872)$ to be $11 \pm 7$ MeV below the $D\bar{D}^{*}$ threshold \cite{Prelovsek:2013cra}. We estimate $X(3872)$ can still be an on-threshold state within our systematic errors. Future $X(3872)$ simulations should be done on larger volumes, to reduce the finite volume systematic errors, which plague bound states in general. Additional scattering levels, that accompany a larger volume, would also provide more phase shift points and thus increase the reliability of the extracted quantities.

\vspace{-0.4cm}
\section{$Z_c^+(3900)$}
\begin{figure}[hbt]
\centering
\vspace{0.7cm}
\includegraphics[width=0.5\textwidth]{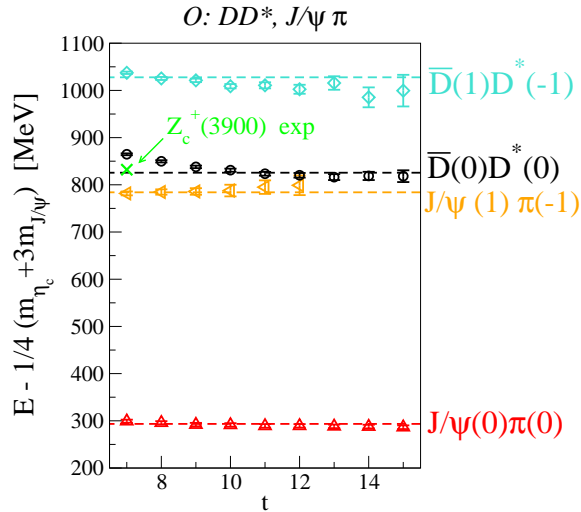} 
\caption{Energy levels in charmonium-like channel with $J^{PC}=1^{+-}$ and $I=1$. We find all the expected scattering states, but no candidate for $Z_{c}^{+}(3900)$.}
\label{Zfig}
\end{figure}

Very recently an interesting state was observed by BESIII in the $J/\psi \pi^+$ invariant mass \cite{Ablikim:2013mio}. The $Z_{c}^{+}(3900)$ is a charmonium-like state, yet it carries charge making it likely to have a $\bar{c}c\bar{d}u$-like structure. Its mass, $m_{Z_{c}}=3899 \pm 6.1$ MeV is very close to the $\bar{D}D^*$ threshold, which suggests that the closeness of the threshold is related to its existence. Experiment has determined its charge conjugation number to be $C=-1$, however parity P and spin J have not been determined yet. Assuming its decay into $J/\psi \pi$ is in s wave we search for $Z^{+}_{c}(3900)$ in the $J^{PC}=1^{+-}$ channel using 3 $J/\psi \pi$ and 3 $\bar{D}D^{*}$ scattering operators as listed in \cite{Prelovsek:2013xba}.\\
The spectrum is presented in Fig. \ref{Zfig}. We observe all expected scattering levels, as they are marked on the right hand side of Fig. \ref{Zfig}, but find no additional energy level \cite{Prelovsek:2013xba} and thus no candidate for $Z_{c}^{+}(3900)$.
It is possible, that we do not find the $Z_{c}^{+}(3900)$ either because the quantum numbers of the state are not $J^{PC}=1^{+-}$, or because our interpolating operator basis is not diverse enough. Future studies should probe also other $J^{PC}$ and include also more exotic types of operators, such as the color triplet ($[qq]^{\bar{3}_c} [\bar{q}\bar{q}]^{3_c}$), color sextet ($[qq]^{6_c} [\bar{q}\bar{q}]^{\bar{6}_c}$) or even the color octet ($[q\bar{q}]^{8_c}[q\bar{q}]^{8_c}$) combination.

\vspace{-0.4cm}
\section{$\chi_{c0}^{'}$}
The $J^{PC}=0^{++}$ charmonium channel has recently become very interesting, as the Particle Data Group \cite{pdg:2012} identified the $\chi_{c0}(2P)$ with $X(3915)$. However some works \cite{Guo:2012tv} do not quite agree with this. In an exploratory study to see what lattice QCD can contribute to this puzzle, we employ 14 interpolating operators:
\begin{align}
&O_{1-9}^{\bar{c}c} = \bar{c} M c (0)\\
&O_1^{DD}= [\bar c \gamma_5 u(0)~\bar u\gamma_5 c(0) +  \{u\to d\} \nonumber\\
&O_2^{DD}\!\!\!\!=[\bar c \gamma_5 \gamma_t u(0)~\bar u\gamma_5 \gamma_t c(0) + \{u\to d\}  \nonumber\\
&O_3^{DD}=\!\!\!\!\!\!\!\!\sum_{e_k=\pm e_{x,y,z}}\!\!\!\! [\bar c \gamma_5 u(e_k)~\bar u\gamma_5 c(-e_k) + \{u\to d\}\nonumber\\
&O_1^{J/\psi \omega}=\sum_{j} ~\bar c \gamma_j c(0)~[\bar u\gamma_j u(0)  + \{u\to d\}\nonumber\\
&O_2^{J/\psi \omega}=\sum_{j} ~\bar c \gamma_j \gamma_t c(0)~[\bar u\gamma_j \gamma_t u(0) + \{u\to d\}\nonumber
\end{align}
Operators $M$ can be found in Table X of \cite{Mohler:2012na}, under the $A_1^{++}$ item. \\
\begin{figure}[hbt]
\vspace{0.5cm}
\centering
\includegraphics[width=0.56\textwidth]{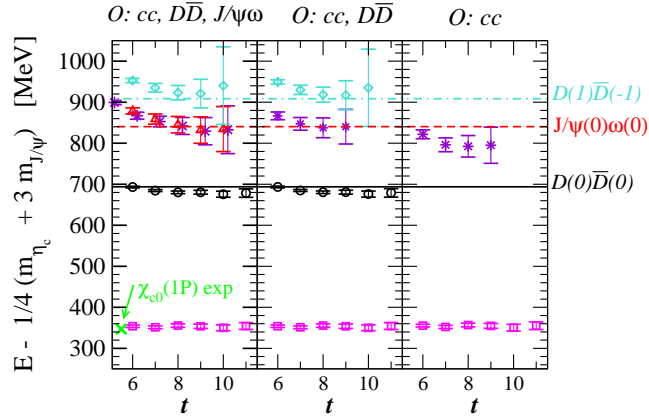} 
\caption{Energy levels for charmonium in the $J^{PC}=0^{++}$ and $I=0$ channel.}
\label{Cfig}
\end{figure}
In Fig. \ref{Cfig} the scattering states are identified and labeled. We find two additional states, where the lowest one (pink squares) is identified with the $\chi_{c0}(1P)$. Both additional states are also present when only $\bar{c}c$ interpolating operators are used. By identifying the $D\bar{D}$ and $J/\psi \omega$ scattering states, it becomes clear that the second additional energy (purple stars) is not a scattering state. The only interpretation left is that it is related to a resonance, call it  $\chi_{c0}^{'}$.\\
As can be seen from the leftmost and middle subplots of Fig. \ref{Cfig} $J/\psi \omega$ interpolating operators have very little effect on the resonance state (purple stars), so we omit them from analysis and assume $D\bar{D}$ elastic scattering.

\begin{figure}[hbt]
\vspace{0.5cm}
\centering
\includegraphics[width=0.5\textwidth]{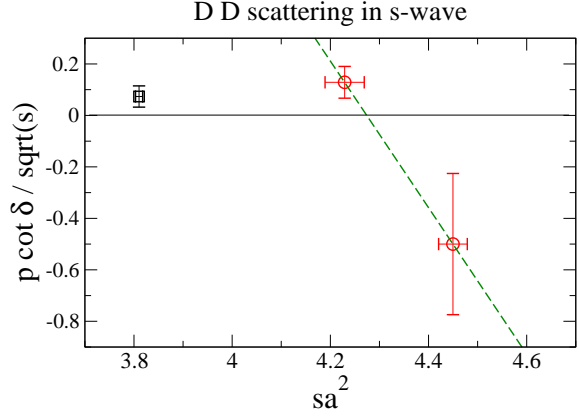} 
\caption{Preliminary phase shift analysis of $D\bar{D}$ scattering in $J^{PC}=0^{++}$ and $I=0$ charmonium channel.}
\label{Cfig2}
\end{figure}

We perform a preliminary phase shift analysis on second, third and fourth level of the middle subplot of Fig. \ref{Cfig}, which is presented in Fig. \ref{Cfig2}. To extract the resonance mass and decay width from the data, we fit a Breit-Wigner curve to two points (red circles in Fig. \ref{Cfig2}):
\vspace{-0.2cm}
\begin{align}
\frac{p\cot \delta}{\sqrt{s}}=\frac{1}{g_{\chi_{c0}^{'}}^2}(m_{\chi_{c0}^{'}}^2-s)\qquad \Gamma[\chi_{c0}'\to \bar DD]\  \equiv g_{\chi_{c0}^{'}}^2~\frac{p}{s}.
\end{align} 
From the fitted results we find the energy difference to be $m_{\chi_{c0}^{'}} - m_{sa}=864 \pm 25$ MeV together with the $\chi_{c0}'\to D\bar D$ coupling $g_{\chi_{c0}^{'}}=0.94 \pm 0.23$ GeV. We add the physical charmonium spin averaged mass to the energy difference, to obtain an estimate for the predicted resonance mass $m_{\chi_{c0}^{'}}^{predict}=3932 \pm 25$ MeV combined with $\Gamma[\chi_{c0}'\to  D\bar{D}]^{predict}=36 {+17\atop -36}$ MeV.\footnote{The error takes into account also variations from the fit-range. } We prompt experiments to look for such state, that  would appear as a peak in the $D\bar D$ invariant mass close to peak related to $\chi_{c2}(2P)$. A possible presence of a a narrow peak in the BaBar and Belle data was pointed out in \cite{Chen:2012wy}. \\
The black square in Fig. \ref{Cfig2} lies significantly away from the one-resonance ($\chi_{c0}'$)  Breit-Wigner fit and indicates increased phase-shift near threshold. This might be related to the broad structure in Belle and BaBar data slightly above $D\bar D$ threshold pointed out by \cite{Guo:2012tv}.  However our lattice results do not support two resonances with two full $2 \pi$ circles in the $D \bar{D}$ phase shift within  the explored energy region. If there was another elastic resonance (for example the one suggested in \cite{Guo:2012tv}) in addition to the $\chi_{c0}^{'}$, then it would appear as another additional state, which we do not observe.\\
The spectrum of the $0^{++}$ charmonium  therefore remains an open question, with exception of $\chi_{c0}(1P)$ ground state. There are experimental and lattice  candidates for states, however their decay channels and widths do not seem to fit together in one picture. We believe that more experimental as well as theoretical studies of this channel should be performed in order to clarify the situation.

\section{Conclusions}
We reported on recent lattice QCD simulations involving several charmonium-like states. In the $J^{PC}=1^{++}$ simulation we found a candidate for $X(3872)$ in $I=0$ and extracted $D\bar{D}^{*}$ scattering length.  We investigated the channel with $J^{PC}=1^{+-}$ and  $I=1$ using $J/\psi \pi$ and $\bar{D}D^{*}$ interpolating operators and found no candidate for the exotic $Z_{c}^{+}(3900)$. We showed preliminary results of the channel with $J^{PC}=0^{++}$ and $I=0$, where we used $\bar{c}c$ as well as $D\bar{D}$ and $J/\psi \omega$ interpolating operators and found a candidate for a resonance, $\chi_{c0}^{'}$, that couples to $D\bar{D}$.

\acknowledgments{We thank Anna Hasenfratz for providing the gauge configurations. We are grateful to C.B. Lang for valuable discussions. The calculations were performed at Jozef Stefan Institute. This work is supported by he Slovenian Research Agency ARRS N1-0020 and by the Austrian Science Fund FWF project I1313-N27. Fermilab is operated by Fermi Research Alliance, LLC under Contract No. De-AC02-07CH11359 with the United States Department of Energy.}

\bibliographystyle{h-physrev4.bst}
\bibliography{refs}

\end{document}